\documentclass[aps,PRB,twocolumn,superscriptaddress,floats]{revtex4}
\usepackage{amssymb}
\usepackage{graphicx}
\usepackage{sidecap}

\begin{document}

\title{Symmetry breaking via orbital-dependent reconstruction of electronic structure in uniaxially strained NaFeAs}

\author{Y. Zhang}
\author{C. He}
\author{Z. R. Ye}
\author{J. Jiang}
\author{F. Chen}
\author{M. Xu}
\author{Q. Q. Ge}
\author{B. P. Xie}
\affiliation{State Key Laboratory of Surface Physics,  Key Laboratory of Micro
and Nano Photonic Structures (MOE), and Department of Physics, Fudan
University, Shanghai 200433, People's Republic of China}

\author{J. Wei}
\author{M. Aeschlimann}

\affiliation{Fachbereich Physik and Research Center OPTIMAS, Technische Universitaet Kaiserslautern, Kaiserslautern, 67663, Germany}

\author{X. Y. Cui}
\author{M. Shi}

\affiliation{Swiss Light Source, Paul-Scherrer Institut, 5232 Villigen, Switzerland}

\author{J. P. Hu}

\affiliation{Department of Physics, Purdue University, West Lafayette, Indiana
47907, USA}

\author{D. L. Feng}\email{dlfeng@fudan.edu.cn}
\affiliation{State Key Laboratory of Surface Physics,  Key Laboratory of Micro
and Nano Photonic Structures (MOE), and Department of Physics, Fudan
University, Shanghai 200433, People's Republic of China}

\begin{abstract}

The superconductivity discovered in iron-pnictides is intimately related to a nematic ground state, where the $C_4$ rotational symmetry is broken via the structural and  magnetic transitions. We here study the nematicity in NaFeAs with the polarization dependent angle-resolved photoemission spectroscopy. A uniaxial strain was applied on the sample to overcome the twinning effect in the low temperature $C_2$-symmetric  state, and obtain a much simpler  electronic structure than that of a twinned sample. We found the electronic structure undergoes an orbital-dependent reconstruction in the nematic state,  primarily involving  the $d_{xy}$- and $d_{yz}$-dominated bands. These bands strongly hybridize with each other, inducing a band splitting, while the $d_{xz}$-dominated bands only exhibit an energy shift without any reconstruction. These findings suggest that  the development of orbital-dependent spin polarization  is  likely the dominant force to drive the nematicity, while  the ferro-orbital ordering between $d_{xz}$ and $d_{yz}$  orbitals can only play a minor role here.
\end{abstract}


\maketitle

\section{Introduction}

The parent compounds of iron-pnictides share a common spin density wave or collinear antiferromagnetic (CAF) state which is characterized by a ferromagnetic (FM) spin alignment along one direction in the  two dimensional rectangular lattice formed by iron sites  and an antiferromagnetic (AFM) spin alignment along the other direction \cite{Neutron1, Neutron2, Neutron3}. Superconductivity in iron-pnictides  emerges when the magnetic order is suppressed through chemical substitution or physical pressure \cite{Phase1, Phase2, Phase3}.  The mechanism causing the CAF order thus may  be intimately related to the superconductivity.

The development of the CAF order is always accompanied
by  a lattice structure distortion, more specifically, a tetragonal-to-orthorhombic structural distortion.  The  transition temperature ($T_S$) at which the lattice distortion takes place either precedes or coincides with the CAF Neel transition temperature ($T_N$) \cite{Neutron1, Neutron2, Neutron3}. The coincidence between the two transitions can be understood from a simple symmetry analysis. Besides breaking magnetic symmetry,  the CAF  state also breaks $C_4$ lattice symmetry of the tetragonal lattice. This lattice symmetry breaking due to the magnetic configuration can be  separately attributed to the development of a nematic electronic order that only  accounts for  the rotation symmetry breaking but not the symmetry breaking of the magnetic space.  Based on the textbook Ginzburg-Laudau approach, the orthorhombic lattice distortion break the same $C_4$ lattice symmetry.  Therefore, the  nematic electronic order and the orthorhombic lattice distortion  are allowed by symmetry to  be  coupled with each other.
Since the observed  lattice distortion in all families of iron-pnictides  is rather small, the electronic nematism  could be dominant.

So far, there are many experimental evidences supporting the dominance of the electronic nematic behavior in iron-pnictides. Transport and neutron scattering measurements found a strong in-plane anisotropy of the resistivity and magnetic excitations \cite{Neutron4,Detwin1,Detwin2,Detwin4}. Angle-resolved photoemission spectroscopy (ARPES), scanning tunneling microscopy, and quantum oscillation studies further revealed a complex electronic structure with C2 rotational symmetry \cite{Arpes1,Arpes2,Arpes3,STM,Quantum1}.  Moreover, the Fermi surface was suggested to be orbital polarized with almost pure $d_{xz}$ orbital character \cite{Arpes4} in the CAF state. The energy splitting of the $d_{xz}$- and $d_{yz}$-dominated bands was further observed far above $T_N$ \cite{Arpes2}.  The observation of such a ferro-orbital ordering has led to a strong debate on the origin of the nematism: whether the electronic nematicity stems from the magnetic or orbital degree of freedom.  Theoretically, some suggest that the nematicity is originated from spin fluctuations \cite{Spin1,Spin2,Spin3}, while others suggest the CAF state comes from a ferro-orbital ordering \cite{Oo1, Oo2, Oo3, Oo4}. Although following symmetry argument, the  electronic nematicities from both origins must be coupled to each other, direct experimental evidence is needed to answer how strongly such an coupling is, and  whether one is dominant over the other.


Previous ARPES studies mainly focus on the so-called ``122'' series of iron pnictides, such as BaFe$_{2-x}$Co$_x$As$_2$ and CaFe$_2$As$_2$. Although the mechanically detwinning methods were used to avoid the complexity due to the twinning effect in the CAF state, the observed electronic structures are still very complicated \cite{Arpes1,Arpes2,Arpes3}. It is thus difficult to obtain a comprehensive understanding of the complex electronic structure and its orbital character in the nematic state. In this paper, we report a systematic ARPES study on the uniaxially strained NaFeAs. The observed electronic structure is much simpler than those of the 122 series. Because of the polarization-sensitivity of orbitals in photoemission experiment, we could selectively probe the bands with different orbital characters. We found that the $C_4$ rotational symmetry of the electronic structure is broken via band reconstruction and the band reconstruction itself  is also strongly orbital-dependent: the $d_{xy}$- and $d_{yz}$-dominated bands open a large hybridization gap at all the band crossings, while the $d_{xz}$-dominated bands only exhibit an energy shift without any reconstruction. Moreover, the observed total occupation of $d_{xz}$ is almost invariant. These results suggests that the ferro-orbital ordering between $d_{xz} $ and $d_{yz}$ can only play a very limited role in driving the electronic nematicity. On the other hand,  the orbital weight redistribution  involving  the $d_{yz}$ and $d_{xy}$ orbitals can only  be explained  through a strong coupling between spin and orbital degrees of freedom far above the Neel transition temperature, which suggests that the nematic spin fluctuations at high temperatures and their coupling with orbital degree of freedom could be the driving force behind both the structural and Neel transitions in iron-pnictides.

\section{Experimental}

\begin{figure}[t]
\centerline{\includegraphics[width=8.7cm]{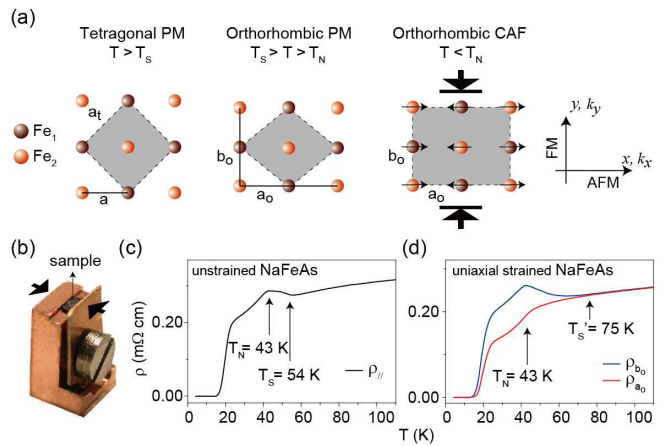}}
\caption{(Color online) (a) Cartoon of the lattice and spin structure in tetragonal PM, orthorhombic PM, and orthorhombic CAF state. The $x$ and $y$ axes are defined along the iron-iron directions. The black arrows show the direction of the uniaxial pressure applied in the mechanical detwinning process. (b) The photograph of the device used to detwin the samples in our experiments.  (c) and (d) The temperature dependent resistivity of unstrained and uniaxially strained NaFeAs, respectively. } \label{BACK}
\end{figure}

NaFeAs single crystals were synthesized by a self-flux
method \cite{HeCheng}. Similar to other parent compounds of iron-pnictides, the high temperature phase ($T~>~T_S$) is in a tetragonal paramagnetic (PM) state, where each unit cell has two irons due to the inequivalent positions of arsenic ions in the FeAs layer [Fig.~\ref{BACK}(a)]. Through the structural transition ($a_o > b_o$), there is an orthorhombic PM state ($T_S~>~T~>~T_N$) with C2 rotational symmetry. The CAF magnetic order shows up when entering the orthorhombic CAF state ($T~<~T_N$) \cite{Neutron1, Neutron2,Neutron3}. The $x$- or $k_x$- axes and $y$- or $k_y$-axes are defined along the AFM and FM directions, respectively. Previous ARPES study on NaFeAs observed a complex electronic structure in the orthorhombic CAF state due to twinned domains of the sample \cite{HeCheng}. Taking the advantage of the mechanically detwinning method \cite{Detwin1,Detwin3}, single domain with sufficiently large size was achieved in our experiment by applying uniaxial strain along Fe-Fe direction [Figs.~\ref{BACK}(a) and \ref{BACK}(b)]. As a result, we could investigate the electronic structure of the orthorhombic CAF state without the complexity of twinning.

The temperature-dependent resistivity of unstrained sample is consistent with the neutron scattering and transport measurements \cite{Neutron3, Resis1}. There are two inflexion points at 54 and 43~K, corresponding to the structural transition temperature ($T_S$) and the Neel transition temperature ($T_N$), respectively [Fig.~\ref{BACK}(c)]. Superconductivity shows up at 20~K, which might be due to the doping effect of the deficient or excess sodium ions. When a uniaxial strain is applied, the in-plane anisotropy of the resistivity ($\rho_{b_o} > \rho_{a_o}$) emerges at about 75~K, much higher than $T_S$ and $T_N$ in the unstrained sample [Fig.~\ref{BACK}(d)]. Similar behavior has been reported in uniaxially strained 122 series \cite{Detwin1,Detwin2,Detwin4}. Recent neutron and X-ray scattering experiments have pointed out that the in-plane anisotropy of the resistivity in BaFe$_2$As$_2$ could be related to the lattice distortion \cite{Recent1, Recent2}. The applied uniaxial strain could smear the structural transition and increase the onset temperature of the lattice distortion. Therefore, we attribute the 75~K observed here to be the structural transition temperature in uniaxially strained samples ($T_S^\prime$).

\begin{figure}[t]
\includegraphics[width=8.7cm]{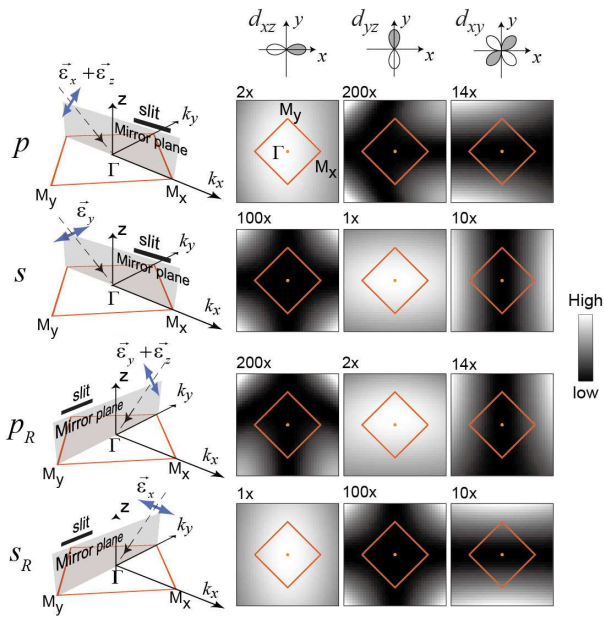}
\caption{(Color online) The experimental setup and matrix element calculations for three $3d$ orbitals. The mirror planes are defined by the analyzer slit and the sample surface normal. The two-dimensional plot of the Brillouin zone of NaFeAs is illustrated by red solid squares.  Note that the photoemission cross-sections are amplified by a factor shown at the up-left corner of the panels. Thus, all the panels could be shown in the same color scales. There is a minor asymmetry in certain distributions caused by the out-of-plane component of the polarization.}\label{SET}
\end{figure}

ARPES studies were performed at the SIS beamline of the Swiss Light Source (SLS). All the data were taken with a Scienta R4000 electron analyzer, the overall energy resolution was 15~meV and the angular resolution was 0.3 degree. The samples were cleaved \textit{in situ}, and measured under ultra-high-vacuum of 5$\times$10$^{-11}$\textit{torr}. The polarization-sensitivity of orbitals in ARPES is a powerful tool to identify the orbital characters of band structure \cite{ZXShenRev, YZhangBaCo}. The photoemission intensity is proportional to the matrix element of the photoemission process ${|M_{f,i}^{\bf{k}}|}$, which can be described by ${|M_{f,i}^{\bf{k}}|\propto{\rm{|}}\langle \phi _f^{\bf{k}}|\bf{\hat{\varepsilon}}\cdot{\bf{r}}|\phi _i^{\bf{k}} \rangle |^2}$, where
$\bf{\hat{\varepsilon}}$ is the unit vector of the electric field of the light, and ${\phi _i^{\bf{k}}}$ is the initial-state wave function. For high kinetic-energy photoelectrons, the final-state wave function ${\phi _f^{\bf{k}}}$ can be approximated by a plane-wave state ${e^{i{\bf{k}\cdot\bf{r}}}}$ with $\bf{k}$ in the mirror plane. By rotating the polarization of incoming photons or the azimuth angle of the sample, four experimental geometries could be achieved named as $p$, $s$, $p_R$, and $s_R$, respectively (Fig.~\ref{SET}). The subscript ``R'' stands for the rotation of the sample. Since the low lying electronic structures at the Fermi energy ($E_F$) of iron-pnictides are commonly believed to be mostly composed of the $d_{xz}$, $d_{yz}$, and $d_{xy}$ orbitals \cite{Orbital}, the matrix element distributions for these three orbitals were calculated in four geometries. As shown in Fig.~\ref{SET}, the matrix element distributions of the $d_{xz}$ and $d_{yz}$ orbitals exhibit a strong polarization dependence through out the first Brillouin zone, reflecting the opposite symmetry of these two orbitals. The matrix element distribution of $d_{xy}$ is suppressed along the direction parallel to the in-plane component of the polarization, due to its odd spatial symmetry with respect to the mirror planes.

\section{The electronic structure in the tetragonal paramagnetic state}

\begin{figure}[t]
\includegraphics[width=8.7cm]{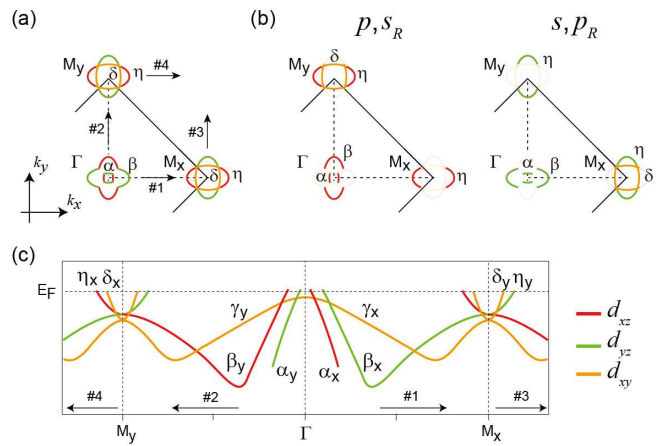}
\caption{ The sketch of the low-lying electronic structure of NaFeAs in the tetragonal PM state. (a) The sketch of the Fermi surface and its orbital character in the folded Brillouin zone for two iron ions per unit cell (solid lines) in the tetragonal PM state. (b) The observable Fermi surfaces in different experimental geometries. (c) The electronic structure of NaFeAs near $E_F$ in the tetragonal PM state.}\label{BAND}
\end{figure}

Based on previous studies \cite{HeCheng, Band}, Fig.~\ref{BAND}(a) sketches the Fermi surface of NaFeAs, which is consisted of two hole pockets ($\alpha$ and $\beta$) around the zone center ($\Gamma$) and two electron pockets ($\eta$ and $\delta$) around the zone corner (M) in the tetragonal PM state. The orbital characters are dominated by the $d_{xz}$, $d_{yz}$ and $d_{xy}$ orbitals \cite{HeCheng2}. The distribution of $d_{xz}$ is equivalent to the $d_{yz}$ orbital after rotating 90 degree, which is protected by the C4 rotational symmetry in the tetragonal PM state. For simplicity, the continuous orbital evolution on the Fermi surface and the orbital mixing on the bands are not illustrated here, which should be finite due to the hybridization of the bands with different orbitals. According to the calculation in section \uppercase\expandafter{\romannumeral2}, we could predict the observable Fermi surfaces in different geometries as shown in Fig.~\ref{BAND}(b). The polarization dependence of the Fermi surface is only sensitive to the in-plane component of the polarization. The low-lying electronic structure of NaFeAs is illustrated in Fig.~\ref{BAND}(c). The bands dispersed along $k_x$ and $k_y$ directions are distinguished with the ``$x$'' and ``$y$'' subscripts here and throughout the paper. As shown by previous study \cite{Orbital}, the $\beta_x$ and $\gamma_x$ bands should connect with the band bottoms of $\eta_y$ and $\delta_y$ at the M point, respectively, which is guaranteed by the continuous evolution of bands in the unfolded Brillouin zone for one iron ion per unit cell.

\begin{figure}[t]
\centerline{\includegraphics[width=8.7cm]{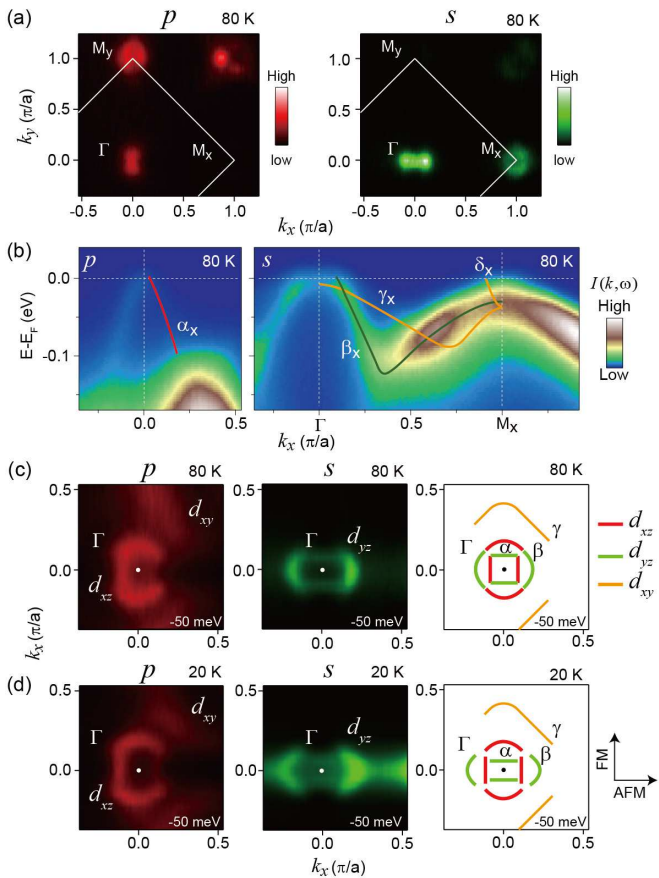}}
\caption{(Color online) The polarization sensitivity of bands with different orbitals. (a) The photoemission intensity maps integrated over a 10~meV energy window at $E_F$ taken at 80~K in the $p$ and $s$ geometries.  (b) The photoemission intensity [$I(k,\omega)$] taken at 80~K in $p$ and $s$ geometries along $\Gamma$-M$_x$ direction. The band dispersions are overlayed on the photoemission intensities with the solid lines. (c) The left and middle panels are the photoemission intensity maps integrated over a 10~meV energy window at 50~meV below $E_F$ taken at 80~K in the $p$ and $s$ geometries, respectively. The observed cross-sections of bands are summarized in the right panel. (d) is same as panel b, but taken at 20~K. All the data are taken with 107~eV photons.}\label{PM}
\end{figure}

\begin{figure*}[t]
\centerline{\includegraphics[width=18cm]{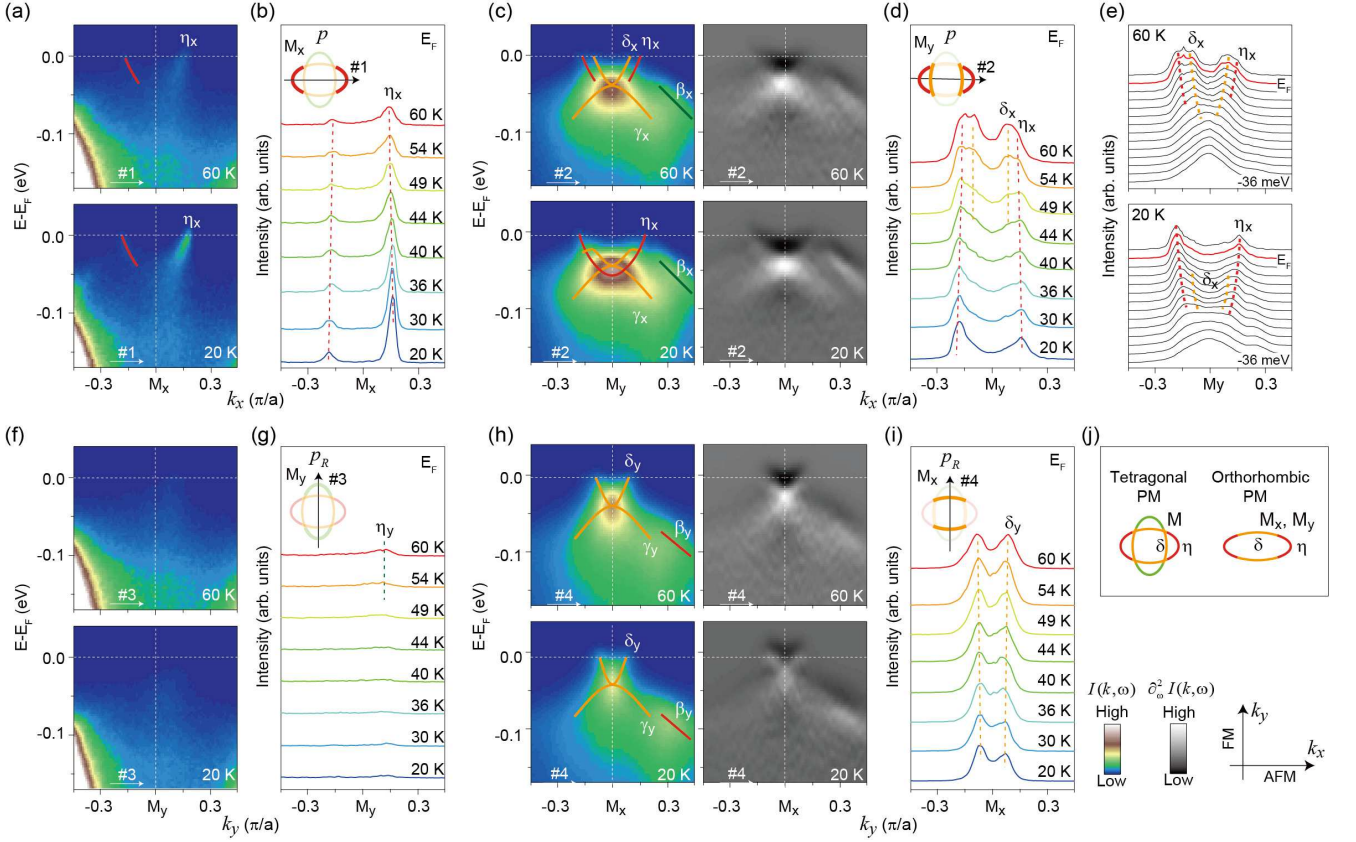}}
\caption{(Color online) The temperature dependent photoemission data of the $\eta$ and $\delta$ bands around the zone corner. (a) The $I(k,\omega)$ taken at 60 and 20~K in the $p$ geometry along Cut \#1. (b) The temperature dependence of the momentum distribution curves (MDCs) at $E_F$ along Cut \#1. The momentum position of Cut \#1 is shown by a black arrow in the inset of panel b. (c) The $I(k,\omega)$ and its second derivative image with respect to energy [$\partial^2 I(k,\omega)/\partial\omega^2$] taken at 60 and 20~K in the $p$ geometry along Cut \#2. (d) The temperature dependence of the MDCs at $E_F$ along Cut \#2. The momentum position of Cut \#2 is shown by a black arrow in the inset of panel d. (e) The MDCs selected between -36~meV to 3~meV for the data in panel a. Each MDCs has been individually normalized by its integrated weight to highlight weak features. (f) and (g) are the same as panel a and b, but taken in the $p_R$ geometry along Cut \#3. (h) and (i) are the same as panel c and d, but taken in the $p_R$ geometry along Cut \#4. (j) The sketch of the electron pockets in the tetragonal PM and orthorhombic PM states. All the data are taken with 118~eV photons. The color scale of $I(k,\omega)$ and $\partial^2 I(k,\omega)/\partial\omega^2$ are shown in the bottom-right inset. These color scales and notations are used throughout the paper. The band dispersions are overlayed on the photoemission intensities with the solid lines, which are determined with the help of the second derivative images and MDCs. } \label{SDW1}
\end{figure*}

To further confirm the orbital character of the electronic structure, the photoemission intensity maps obtained in $p$ and $s$ geometries are compared in Fig.~\ref{PM}(a). Around $\Gamma$, the photoemission intensity distribution rotates 90 degree from the $p$ to $s$ geometry, which is consistent with the predicted distribution of the $d_{xz}$ and $d_{yz}$ orbitals on the $\alpha$ and $\beta$ hole pockets. The electron pockets are observed either around $M_y$ in the $p$ geometry or around $M_x$ in the $s$ geometry, which is corresponding to the $\delta$ electron pockets dominated by the $d_{xy}$ orbital. We did not observe the $\eta$ electron pocket here, which could be due to particular photon energy or experimental setup, since the outer $\eta$ electron pockets could be observed in previous studies \cite{HeCheng} or with 118~eV photons as shown later in Fig.~\ref{SDW1}.

\begin{figure*}
\centerline{\includegraphics[width=18cm]{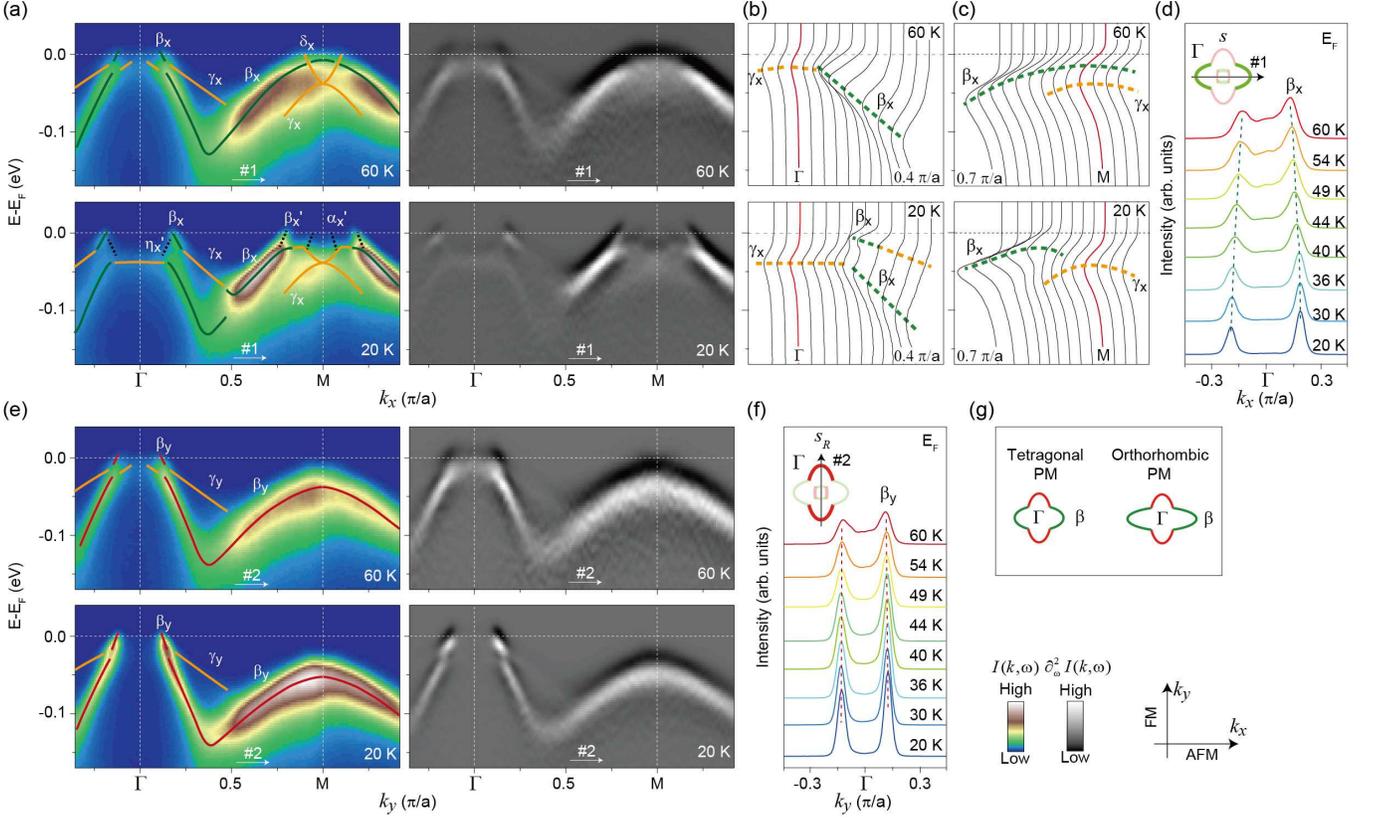}}
\caption{(Color online) The temperature dependent photoemission data of the $\beta$ and $\gamma$ bands. (a) The $I(k,\omega)$ and $\partial^2 I(k,\omega)/\partial\omega^2$ taken at 60 and 20~K in the $s$ geometry along Cut \#1. (b) and (c) The Energy Distribution Curves (EDCs) selected from the data in panel a around $\Gamma$ and M, respectively. (d) The temperature dependence of the MDCs at $E_F$ along Cut \#1. The momentum position of Cut \#1 is shown by a black arrow in the inset of panel d. (e) and (f) are the same as panel a and d, but taken in the $s_R$ geometry along Cut \#2. (g) The sketch of the $\beta$ hole pocket in the tetragonal PM and orthorhombic PM states. All the data are taken with 118~eV photons. } \label{SDW2}
\end{figure*}

For the low-lying electronic structure, the bands dominated by $d_{xz}$, $d_{yz}$, and $d_{xy}$ could be selectively probed in different geometries [Fig.~\ref{PM}(b)]. The $\alpha_x$ band appears in the $p$ geometry, while the $\beta_x$ band could be observed in the $s$ geometry. The $\gamma_x$ and $\delta_x$ bands show weak photoemission intensities due to the small matrix element of the $d_{xy}$ orbital. These two bands could be observed more clearly in Figs.~\ref{SDW1} and ~\ref{SDW2}. We further investigated the photoemission intensity maps at 50~meV below $E_F$ [Fig.~\ref{PM}(c)].  Two ellipse-like features observed in $p$ and $s$ geometries are originated from the $d_{xz}$ and $d_{yz}$ orbital, respectively. They intertwined with each other forming the cross-sections of the $\alpha$ and $\beta$ hole pockets. The cross-sections of $\gamma$ is suppressed along the $k_x$ direction in the $p$ geometry, which is consistent with the matrix element distribution of $d_{xy}$. When the sample is cooled down to 20~K, the symmetry breaking of electronic structure is notable on the cross-sections of the $\alpha$ and $\beta$ hole pockets  [Fig.~\ref{PM}(d)]. Both pockets extend along the AFM direction. The $d_{yz}$ portions show more anisotropy than the $d_{xz}$ portions indicating an orbital-dependent reconstruction of the electronic structure.

\begin{figure*}
\centerline{\includegraphics[width=18cm]{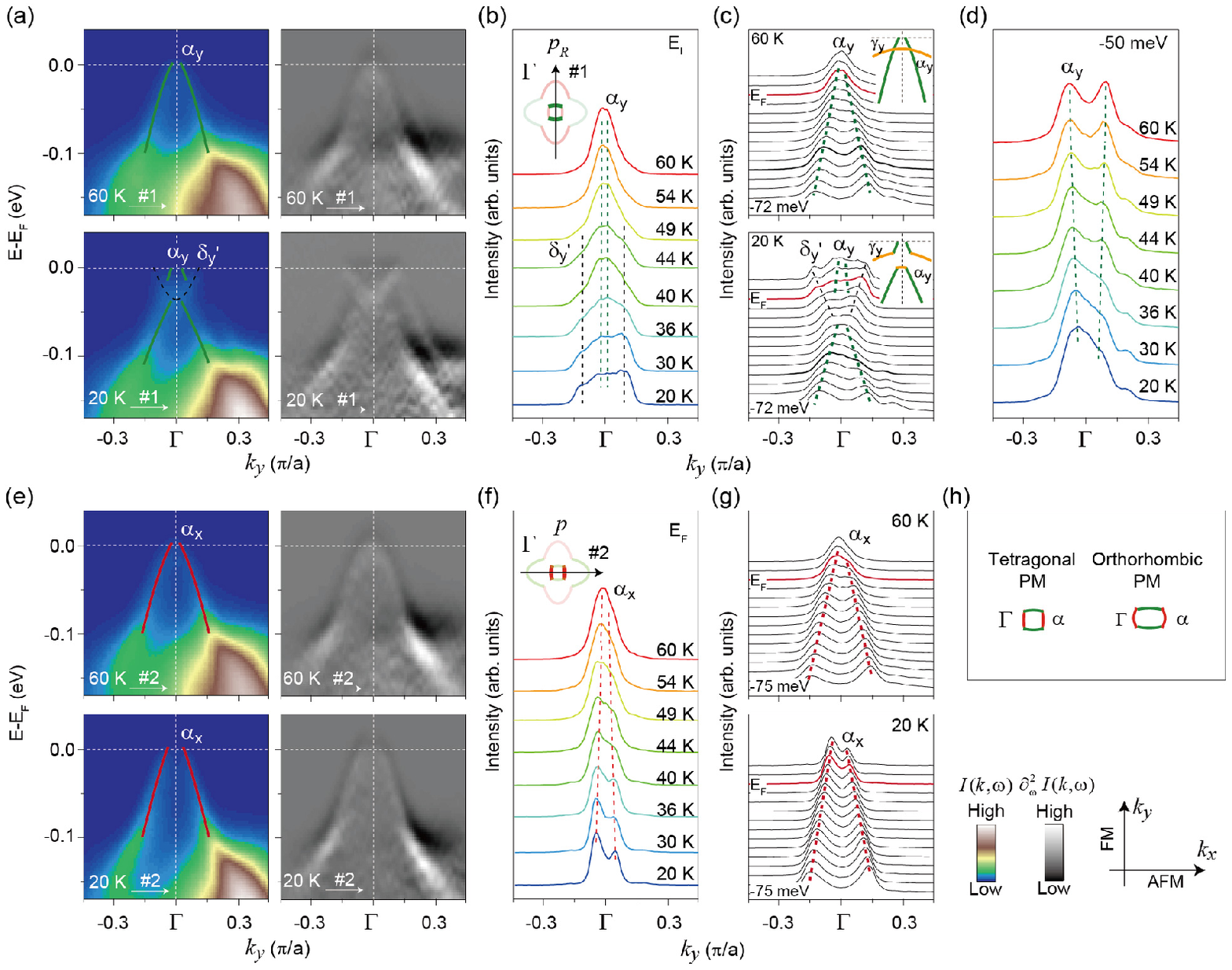}}
\caption{(Color online) The temperature dependent photoemission data of the $\alpha$ band around the zone center. (a) The $I(k,\omega)$ and $\partial^2 I(k,\omega)/\partial\omega^2$ taken at 60 and 20~K in the $p_R$ geometry along Cut \#1. (b) The temperature dependence of the MDCs at $E_F$ along Cut \#1. The momentum position of Cut \#1 is shown by a black arrow in the inset of panel b. (c) The MDCs selected between -72~meV to 6~meV for the data in panel a. Each MDCs has been individually normalized by its integrated weight to highlight weak features. (d) The temperature dependence of the MDCs at 50~meV below $E_F$ along Cut \#1.  (e), (f), and (g) are the same as panel a, b, and c, but taken in the $p$ geometry along Cut \#2. (h) The sketch of the $\alpha$ hole pocket in the tetragonal PM and orthorhombic PM states. All the data are taken with 118~eV photons. }\label{SDW3}
\end{figure*}

\section{The electronic structure of the nematic state}

As shown in section \uppercase\expandafter{\romannumeral3}, although the simple matrix element calculation is not quantitative, there is a good qualitative agreement between the prediction and our experimental results. Therefore, NaFeAs provides an ideal opportunity for us to study the temperature dependence of $d_{xz}$, $d_{yz}$, and $d_{xy}$ separately along both AFM and FM directions. We will focus on the temperature dependence of the bands that make the electron pockets first.

\subsection{The temperature dependence of the $\eta$ and $\delta$ bands}

There are two electron pockets $\eta$ and $\delta$ around the zone corner. The $\eta_x$ bands could be clearly identified crossing $E_F$ along the $k_x$ direction in the $p$ geometry [Fig.~\ref{SDW1}(a)]. Their Fermi crossings move outward continuously as the temperature decreases [Fig.~\ref{SDW1}(b)]. The momentum shift of the $\eta_x$ Fermi crossing is about 0.02~$\pi$/a from 60 to 20~K. For the $\delta$ electron pocket, we took the data along the momentum cut around $M_y$ in the $p$ geometry in order to enhance its photoemission intensity. As shown in Figs.~\ref{SDW1}(c)-\ref{SDW1}(e), when the temperature decreases, the MDCs peak intensity of $\delta_x$ suppresses near $E_F$ promptly, indicating an energy gap opening for the $\delta_x$ band along the $k_x$ direction. Note that, the band dispersions along the $k_x$ direction should be equivalent for $M_x$ and $M_y$, which is protected by the translation symmetry.

Along the $k_y$ direction, the photoemission intensity of $\eta_y$ is too weak to be distinguished [Fig.~\ref{SDW1}(f)]. As will be discussed later, this is because the $\eta_y$ band shifts up above $E_F$. The tiny MDCs peaks observed in Fig.~\ref{SDW1}(g) might be due to the minor twinned domains of the sample. For the $\delta_y$ portions of the inner electron pocket, weak temperature dependence was observed [Figs.~\ref{SDW1}(h) and \ref{SDW1}(i)]. The $\delta_y$ Fermi crossings slightly shrink with a momentum shift of about 0.02~$\pi$/a from 60 to 20~K.

The in-plane anisotropy of the electron pockets preserves at the temperature as high as 60~K, which is far above the structural and magnetic transition temperatures in unstrained sample. This suggests that the anisotropic electronic structure observed here might be closely related to the strain and thus lattice distortion, whose onset temperature ($T_S^\prime$) increases due to the uniaxial strain (More evidences in Fig.~\ref{SUM1}). Therefore, in the orthorhombic PM state above $T_N$, the electron pockets reconstruct significantly as shown in Fig.~\ref{SDW1}(j). Two portions vanish in the the orthorhombic PM state and the remaining electron pocket extends along the $k_x$ direction.

\subsection{The temperature dependence of the $\alpha$, $\beta$, and $\gamma$ bands}

To investigate the anisotropy of the $\beta$ and $\gamma$ bands, temperature dependent data taken in the $s$ geometry along the $k_x$ direction are shown in Fig.~\ref{SDW2}(a). Remarkable band reconstruction was observed. Near $\Gamma$, the $\beta_x$ and $\gamma_x$ bands open a large hybridization gap at 20~K. As a result, the Fermi crossings of $\beta_x$ are pushed outward from $\Gamma$, and the band top of $\gamma$ shifts downward to about 40~meV below $E_F$ [Figs.~\ref{SDW2}(a), \ref{SDW2}(b), and \ref{SDW2}(d)]. The $\beta_x$ band disperses towards $E_F$ around M forming a hole-like band. It hybridizes with the $\delta_x$ electron pocket and opens an energy gap at $E_F$ upon lowing the temperature. This is consistent with the gap opening of $\delta_x$ observed in Figs.~\ref{SDW1}(c)-\ref{SDW1}(e). Moreover, the band shift and hybridization push the $\beta_x$ band above $E_F$ around M. Considering the fact that the $\beta_x$ band should connect with the band bottom of $\eta_y$ [Fig.~\ref{BAND}(c)], we could further confirm the observation in Fig.~\ref{SDW1}(f) that the $\eta_y$ portions of the outer electron pocket shift up above $E_F$ in the orthorhombic PM state. Note that, the long-range magnetic order could halve the Brillouin zone by folding the bands between $\Gamma$ and M. We thus attributed $\eta_x^{\prime}$ near $\Gamma$ together with $\beta_x^{\prime}$ and $\alpha_x^{\prime}$ around M to be the folded bands in the magnetic state.

In contrast to the complex behavior along $k_x$, the $\beta$ band does not reconstruct along the $k_y$ direction [Fig.~\ref{SDW2}(e)]. The Fermi crossings of $\beta_y$ are almost temperature independent as shown in Fig.~\ref{SDW2}(f). Around the M point, the $\beta_y$ band shifts downward to about -50~meV at the lowest temperature. Since $\beta_y$ should connect with the band bottom of $\eta_x$ [Fig.~\ref{BAND}(d)], such an energy shift of $\beta_y$ is consistent with the enlargement of $\eta_x$ observed in Fig.~\ref{SDW1}(a). We summarized the reconstruction of $\beta$ in Fig.~\ref{SDW2}(g). The $\beta$ hole pocket extend along the AFM direction in the orthorhombic PM state.

The reconstruction of the inner $\alpha$ hole pocket is shown in Fig.~\ref{SDW3}. Along the $k_y$ direction, the $\alpha_y$ band split into two parts indicating possible hybridization with other bands [Figs.~\ref{SDW3}(a) and \ref{SDW3}(c)].  The hybridization with $\delta_y^\prime$ could be first excluded. According to the temperature dependent data, while band shift of the lower part of $\alpha_y$ is notable at 60~K [Fig.~\ref{SDW3}(d)], the folding of $\delta_y$ occurs at a much lower temperature of about 49~K [Fig.~\ref{SDW3}(b)]. Moreover, no hybridization was found for the $\delta_y$ band in Fig.~\ref{SDW1}(h). Therefore, the $\alpha_y$ band should hybridize strongly with the $\gamma_y$ band in the orthorhombic PM state, as shown in the inset of Fig.~\ref{SDW3}(c). In that case, a large proportion of the $d_{yz}$ orbital on $\alpha_y$ would mixed into $\gamma_y$ due to such hybridization. Consequently, we note that the photoemission intensity of $\gamma_y$ is suppressed around the $\Gamma$ point at 20~K in Fig.~\ref{SDW2}(e), since the $d_{yz}$ orbital could not be observed in the $s_R$ geometry. Along the $k_x$ direction, the $\alpha_x$ bands show weak temperature dependence. Their Fermi crossings slightly move outward about 0.02~$\pi$/a from 60 to 20~K [Figs.~\ref{SDW3}(e)-\ref{SDW3}(g)]. Therefore, similar to the $\beta$ hole pocket, the $\alpha$ hole pocket extends along the AFM direction in orthorhombic PM state as shown in Fig.~\ref{SDW3}(h).

Our previous APRES study on unstrained NaFeAs has found that the short range magnetic order emerges just below the structural transition temperature of 54~K, which causes a band folding \cite{HeCheng}. However, when the structural transition is enhanced to  75~K  here by the uniaxial strain, we still observe  the folding of $\delta_y$ bands below $\sim$49~K [Fig.~\ref{SDW3}(b)]. Therefore, the short range magnetic order observed above $T_N$ in previous study should be related to the fluctuation of the long range magnetic order, instead of a direct correlation with the structural transition.

\section{Discussion}

\begin{figure}
\centerline{\includegraphics[width=8.7cm]{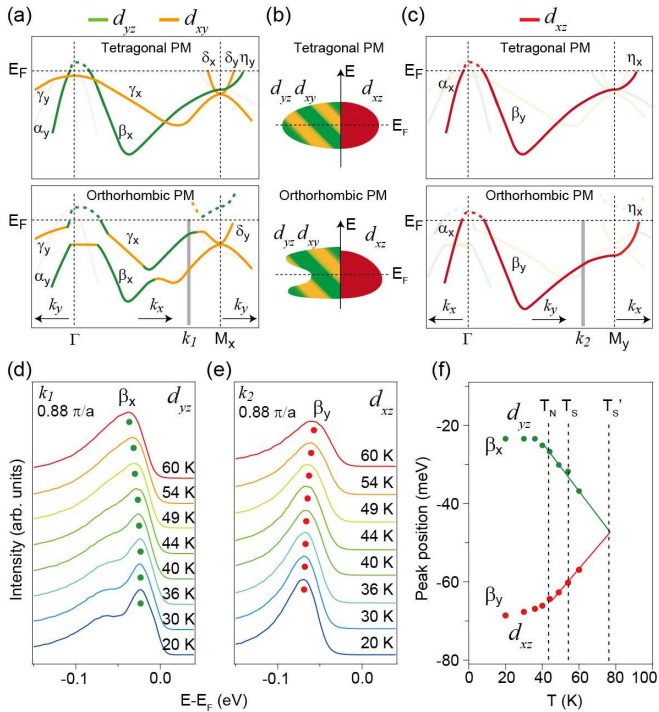}}
\caption{(Color
online) The orbital-dependent reconstruction of the electronic structure. (a) The low-lying electronic structure in the tetragonal PM and orthorhombic PM state respectively, where only the $d_{yz}$- and $d_{xy}$-dominated bands are highlighted. The band dispersions above $E_F$ are shown considering the continuous dispersion of the bands with certain orbitals.  (b) Cartoon shows the orbital weight distributions near $E_F$ in the tetragonal PM and orthorhombic PM states. (c) The low-lying electronic structure in the tetragonal PM and orthorhombic PM state respectively, where only the $d_{xz}$-dominated  bands  are highlighted. (d) The temperature dependence of the EDCs at $k_1$ as indicated by the gray line in panel a. (e) The temperature dependence of the EDCs at $k_2$ as indicated by the gray line in panel c. (f) The peak positions of the $\beta_x$ and $\beta_y$ bands as functions of temperature.} \label{SUM1}
\end{figure}

To further investigate the orbital-dependent behaviors of the reconstruction in the orthorhombic PM state, the bands dominated by the $d_{yz}$ and $d_{xy}$ orbitals are highlighted in Fig.~\ref{SUM1}(a). The most remarkable feature found here is the hybridization gap opening at almost all band crossings. As a result, the bands split into the upper and lower bands around $\Gamma$ and open an energy gap at $E_F$ around M. Such reconstruction could redistribute the orbital weight of $d_{yz}$ and $d_{xy}$, inducing a partial gap at $E_F$ [Fig.~\ref{SUM1}(b)]. On the contrary, the $d_{xz}$-dominated bands exhibit no remarkable reconstruction, except an energy shift around M [Fig.~\ref{SUM1}(c)]. The orbital weight redistribution is negligible for the $d_{xz}$ orbital. Since the energy shift of the $\beta$ band is pronounced near the M point, we use this energy shift as a parameter to describe the reconstruction of the electronic structure. If the C4 rotational symmetry of the electronic structure is preserved, the $\beta$ band dominated by the $d_{xz}$ and $d_{yz}$ orbitals should degenerate along the $k_x$ and $k_y$ directions. As shown in Figs.~\ref{SUM1}(d)-\ref{SUM1}(f), when the temperature decreases, the energy splitting between $\beta_x$ and $\beta_y$ bands enlarge linearly and almost saturate at $T_N$. The onset temperature of such splitting could be estimated to be around 75~K, where the in-plane anisotropy of resistivity emerges ($T_S^\prime$). This result is consistent with previous studies on the ``122" series of iron pnictides \cite{Arpes2}, and suggests that the significant orbital weight redistribution occurs below the structural transition temperature. Moreover, the smooth evolution of the electronic structure reconstruction into the CAF state indicates the both the magnetic and structural transitions share the same driving force \cite{HeCheng}.

\begin{figure}[t]
\centerline{\includegraphics[width=8.7cm]{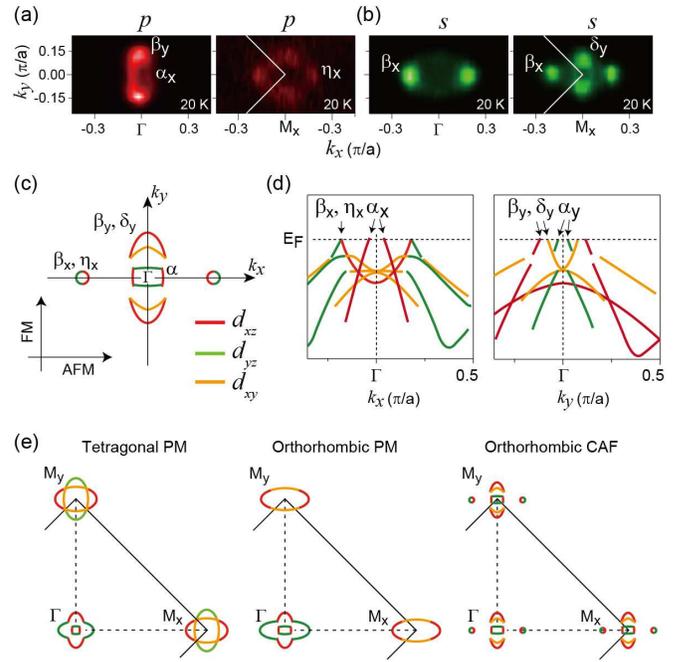}}
\caption{(Color online) The temperature dependence of the Fermi surface morphology. (a) The photoemission intensity maps in the orthorhombic CAF state taken around $\Gamma$ and M in the $p$ geometry. (b) is the same as panel a, but taken in the $s$ geometry. The intensity was integrated over a 10~meV energy window at $E_F$. (c) The Fermi surface morphology in the orthorhombic CAF state. (d) The band structure along $k_x$ and $k_y$ directions in the orthorhombic CAF state. (e) The summary of Fermi surfaces in the tetragonal PM, orthorhombic PM, and orthorhombic CAF state respectively. } \label{SUM2}
\end{figure}

As proposed by many theories, the orbital weight redistribution is important to understand the nematic transitions in iron-pnictides \cite{Oo2,Oo3,Oo4,Oo5,Oo6,Oo7,Oo8}. Our results thus provide several important quantitative tests for these theoretical descriptions. We could first check the inequivalent occupation between the $d_{xz}$ and $d_{yz}$ orbitals, which is considered to be important for the formation of ferro-orbital ordering \cite{Oo2,Oo3,Oo4}. The occupation change of the $d_{xz}$ orbital could be calculated from the size increase of the $\eta_x$ electron pocket subtracted by the size increase of the $\alpha_x$ hole pocket. The momentum shift of the $\eta_x$ Fermi crossings is about 0.02 $\pi$/a from 60~K to 20~K. Considering the band shift above 60~K [Fig.~\ref{SUM1}(f)], this momentum shift might be doubled to 0.04 $\pi$/a corresponding to a maximum occupation change of about 0.04 electron per iron. Moreover, the Fermi momenta variations are almost the same for the enlargement of the $\eta_x$ electron pocket and the $\alpha_x$ hole pocket. Therefore, \emph{the occupation change of the $d_{xz}$ orbital should be infinitesimal}, which suggests that the ferro-orbital ordering only involving $d_{xz}$ and $d_{yz}$
plays a minor role in driving the nematic transitions here \cite{Oo2}. Note that, although the occupation of $d_{xz}$ is almost invariant, the orbital polarization could be strong only at the low energy. Since the orbital weight of $d_{yz}$ and $d_{xy}$ is partially gaped away from $E_F$, the Fermi surface could exhibit more $d_{xz}$ orbital character in orthodromic state \cite{Arpes4,Oo8}.

More intriguingly, instead of the pure ferro-orbital ordering between $d_{xz}$ and $d_{yz}$ orbitals, the reconstruction of the electronic structure involves mostly the $d_{yz}$ and $d_{xy}$ orbitals. These two orbitals are  found numerically to contribute much larger magnetic moment than $d_{xz}$ in the  magnetic state \cite{Oo3,Oo5,Oo6,Oo7,Oo8}. Therefore, the orbital weight redistribution observed here below $T_S^\prime$ could be related to the spin polarization of the $d_{yz}$ and $d_{xy}$ orbitals. In this scenario, the strong coupling between spin fluctuation and orbital degree of freedom could induce an orbital-dependent spin polarization, which further reinforces the spin fluctuation and the anisotropy of electronic structure far above the Neel transition \cite{Oo5,Oo6, Oo7,Oo8}. Such a positive feedback process can be accelerated by the an uniaxial strain, which helps to explain the dramatic enhancement of  $T_S$ by the rather small strain applied in our experiment \cite{Detwin1}. Consistently, the enhancement of spin fluctuation could be observed under the structural transition \cite{NMR}. Our results therefore suggest that the nematic spin fluctuations at high temperatures and their coupling with orbital degree of freedom could be the driving force behind both the structural and Neel transitions in iron-pnictides.

In the magnetic state, the electronic structure is  folded between $\Gamma$ and M due to the long-range magnetic order. According to the photoemission intensity maps in $s$ and $p$ geometries [Figs.~\ref{SUM2}(a) and \ref{SUM2}(b)], the Fermi surface morphology and band structure in the magnetic state are shown in Figs.~\ref{SUM2}(c) and \ref{SUM2}(d). The electron pocket folds from M to $\Gamma$ and then hybridizes with the $\beta$ hole pocket. As a result, four small pockets are formed. There are two bright spots along AFM direction dominated by the $d_{yz}$ and $d_{xz}$ orbitals. Since they are formed by two bands with opposite symmetries [Figs.~\ref{SUM2}(c) and \ref{SUM2}(d)], these two small pockets are intensively discussed as Dirac points in 122 series \cite{Arpes2,Dirac1,Dirac2}. The other two pockets are formed by the $\delta_y$ and $\beta_y$ bands along the FM direction [Figs.~\ref{SUM2}(c) and \ref{SUM2}(d)].

We summarized the reconstruction of Fermi surface in Fig.~\ref{SUM2}(e). The C4 rotational symmetry of the Fermi surface is broken under the structural transition. The anisotropic Fermi surface observed in the the orthorhombic PM state could be responsible for the in-plane anisotropic transport behaviors above $T_N$ \cite{Detwin1, Detwin2, Detwin4, Optic1}. All the pockets show anisotropic shape with an extension along $k_x$ direction. Especially, the transport properties contributed by the electron pockets could be highly anisotropic, since two portions of electron pockets are gaped away from $E_F$ in the orthorhombic PM state. When entering the orthorhombic CAF state, the Fermi pockets are folded between $\Gamma$ and M. The Fermi surface is thus characterized by four distinct pockets originated from the hybridization between the hole and electron pockets. These features  qualitatively agree with previous studies in the 122-series \cite{Arpes1, Arpes2, Oo6, Quantum1}, which highlight the unified nature of the nematic magnetic state in iron-pnictides. Furthermore, the Fermi crossings of $\beta$, $\delta_y$, and $\eta_x$ are sensitive to the structures, $k_z$, and doping levels. Therefore, the topology of these four pockets could be variable in different compounds. For example, the pockets distributed along FM direction are hole-like in NaFeAs, but electron-like in 122 series \cite{Arpes1, Arpes2, Oo6, Quantum1}. In that case, the topology change of these distinct features could be responsible for the non-monotonic doping dependence of the in-plane resistivity anisotropy observed in 122 series \cite{Detwin1,Detwin2,Detwin4,DetwinT}.


\section{Summary}

To summarize, our polarization-dependent ARPES study on uniaxially strained NaFeAs has clearly demonstrate the details of the electronic structure reconstruction in the nematic state. We found that the rotational symmetry breaking of electronic structure is closely related to the in-plane anisotropy of the resistivity, whose onset temperature is far above the Neel transition temperature. More intriguingly, such a symmetry breaking originates form the reconstruction of the electronic structure, and exhibits orbital-dependent behaviors. The $d_{yz}$- and $d_{xy}$-dominated bands split near $\Gamma$ and open an energy gap around $M$, while the $d_{xz}$-dominated bands only exhibit an energy shift without any reconstruction. Our observations exclude the theories based on the ferro-orbital ordering only involving the $d_{xz}$ and $d_{yz}$ orbitals, since the occupation of $d_{xz}$ is almost temperature independent. Instead, the strong coupling between the nematic spin fluctuations and the orbital degree of freedom are suggested to drive the nematicity. Such a coupling causes an orbital-dependent spin polarization, and thus the development of finite magnetic moments of the $d_{yz}$ and $d_{xy}$ orbitals as manifested in their weight redistributions. Our findings provide the experimental foundation for a microscopic understanding of the nematic state of iron-pnictides.

\section{Acknowledgments}

We thank Dr. Chi-Cheng Lee for helpful discussions, and Dr. Dong-hui Lu. and Dr. Changyoung Kim for sharing experiences of the mechanically detwinning method. This work is supported in part by the National Science Foundation of China, Ministry of Education of China, Science and Technology Committee of Shanghai Municipal, and National Basic Research Program of China (973 Program) under the grant Nos. 2011CB921802, 2011CBA00112, and 2012CB921400.

\end{document}